\documentclass[12pt]{article} \textheight 8.5in \textwidth 6.5in
\usepackage{amsmath,amsfonts,latexsym,amssymb}

\oddsidemargin 0in \topmargin -.35in \title{Quantum Field Theory
Is Not Merely Quantum Mechanics Applied to Low Energy Effective
Degrees of Freedom} 
\author{Stefan Hollands\thanks{\tt stefan@gr.uchicago.edu}
and Robert M. Wald\thanks{\tt rmwa@midway.uchicago.edu} \\ \it Enrico
Fermi Institute and Department of Physics \\ \it University of Chicago
\\ \it 5640 S.~Ellis Avenue, Chicago, IL~60637, USA}
\begin{document}
\date{}
\maketitle
\begin{abstract}

It is commonly assumed that quantum field theory arises by applying
ordinary quantum mechanics to the low energy effective degrees of
freedom of a more fundamental theory defined at
ultra-high-energy/short-wavelength scales. We shall argue here that,
even for free quantum fields, there are holistic aspects of quantum field
theory that cannot be properly understood in this
manner. Specifically, the ``subtractions'' needed to define nonlinear
polynomial functions of a free quantum field in curved spacetime
are quite simple and
natural from the quantum field theoretic point of view, but are at best
extremely ad hoc and unnatural if viewed as independent
renormalizations of individual modes
of the field. We illustrate this point by contrasting the analysis of
the Casimir effect, the renormalization of the stress-energy tensor in
time-dependent spacetimes, and anomalies from the point of quantum
field theory and from the point of view of quantum mechanics applied
to the independent low energy modes of the field. Some implications
for the cosmological constant problem are discussed.

\end{abstract}

Quantum field theory provides an excellent description of all
phenomena observed in nature, at least down to the distance scales
probed by present accelerators. Nevertheless, there is good reason to
expect that it will break down at some distance scale $l_0$
(presumably of order of the Planck length) and be replaced by a more
fundamental theory. The most compelling reason to expect such a
breakdown comes from the quantization of gravity: Although a
mathematically rigorous formulation of quantum field theory on a
classical gravitational background can be given, it does not
seem possible to formulate a quantum theory of the spacetime metric
itself within the conventional framework of quantum field theory.

Even if quantum field theory does not provide a fundamental
description of nature, one can attempt to understand its success in
describing low energy phenomena in much the same way as one can
understand why the continuum theory of elasticity is successful in
describing the long wavelength excitations of a crystal. In the case
of a crystal, the continuum theory clearly breaks down at the scale of
the lattice spacing. Nevertheless, starting from the fundamental
lattice theory, one can derive an effective continuum theory that
provides an accurate description of all aspects of the long wavelength
degrees of freedom.

It is widely believed that quantum field theory similarly arises as an
effective field theory from a more fundamental theory defined at
ultra-high-energy/short-wavelength scales. We do not disagree with
this viewpoint and, indeed, this viewpoint has led to many fruitful
insights into the nature of quantum field theory.  However, there is
an associated view, which is implicit in many discussions, that
quantum field theory is nothing more than quantum mechanics applied to
the low energy effective degrees of freedom of the fundamental theory,
in much the same way as elasticity theory is just classical (or
quantum) mechanics applied to the effective degrees of freedom
describing low energy excitations of a crystal. 
The main purpose of
this paper is to point out that, even for a free quantum field, there
are some fundamental features of quantum field theory that cannot be
properly explained in this associated viewpoint.

These fundamental features concern the holistic nature of renormalization
theory. For simplicity and definiteness, we focus our discussion on a free,
massless, Klein-Gordon scalar quantum field
\begin{equation}
\nabla^a \nabla_a \varphi = 0 
\label{kg}
\end{equation}
in curved spacetime. (Similar arguments could be made for any other linear
or nonlinear quantum field theory.)
As is well known,
infinities arise in the calculation of
any field quantity $\Phi$ corresponding to 
a nonlinear polynomial expression
in $\varphi$ and its derivatives. Therefore, ``subtractions'' must
be done to give $\Phi$ a well defined, finite meaning. 
From the quantum
field theoretic viewpoint, it is crucial that $\Phi$
be defined in a local and covariant manner
\cite{hw}, \cite{bfv}. The requirement that the ``subtraction'' be done in a
local and covariant manner greatly constrains the
renormalization procedure. In particular, as proven in \cite{hw}, it
reduces the ambiguities in any polynomial quantity
$\Phi$ to at most a finite number of parameters.

Let us now compare this situation with the picture obtained if we
decompose the quantum field $\varphi$ into modes, and view these modes
as independent degrees of freedom. First, we divide these modes into
``(ultra-)high-energy/short-wavelength'' modes ($\omega \gtrsim 1/l_0$)
and ``low energy modes'' ($\omega \lesssim 1/l_0$), where $l_0$
denotes the physical length scale at which quantum field theory breaks
down. It would be reasonable to assume $l_0$ to be of order the Planck
scale, $l_0 \sim (G \hbar/c^3)^{1/2}$, but we need not make any such
assumption here. We do not know how to accurately describe the
``ultra-high-energy/short-wavelength'' modes using the presently known
laws of physics, but we assume that these modes contribute negligibly
to $\Phi$. On the other hand, we assume that the degrees of freedom
corresponding to the low energy modes are described by ordinary
quantum mechanics. As is well known, the sum of these low energy mode
contributions to any given nonlinear polynomial $\Phi$ is absolutely
enormous (see below). Thus, subtractions are needed to
obtain reasonable renormalized field quantities. In this respect, the
situation with regard to defining $\Phi$ by such a truncated mode sum 
calculation is neither better nor worse than the above field
theoretic calculation; some renormalization is needed in both
cases. However, there is an important difference here in how the
renormalization is to be carried out. The decomposition of $\varphi$
into modes representing its individual degrees of freedom is
necessarily global in character. From the point of view of an
individual mode, there is therefore no natural way of enforcing the
requirement that the subtractions be done in such a way that the
resulting $\Phi$ is defined in a local and covariant manner. In other
words, the local and covariant character of $\Phi$ is a property that
depends on the {\em sum} of all of the renormalized modes. An
individual mode will have no way of knowing whether its own
subtraction is correct unless it ``knows'' how the subtractions are being
done for all other modes. Of course, one can make use of the enormous
available freedom to make arbitrary subtractions to cook up mode-by-mode
renormalization schemes that, by construction, reproduce the field
theoretic renormalization prescription. However, as we shall illustrate
below, in curved spacetime, these schemes are necessarily so ad hoc and
unnatural---far worse than the familiar ``vacuum subtractions'' of Minkowski
spacetime---that it is very difficult to imagine that they could have any
validity. The field theoretic renormalization prescription makes sense
only from a holistic point of view, not from the point of view of 
individual modes. In this sense, even a free quantum field is much more
than the sum of its dynamically independent parts, and quantum field theory
is much more than merely quantum mechanics applied to the individual
low energy degrees of freedom of the field.

The holistic nature of renormalization in quantum field theory is
disguised in Minkowski spacetime on account of the fact that the
locally and covariantly constructed Hadamard distribution $H(x,x')$
that enters the quantum field theoretic renormalization prescription
\cite{hw} for $\Phi$ happens to equal the expection value of $\varphi(x)
\varphi(x')$ in the Minkowski vacuum state $|0 \rangle$. As a consequence, in
Minkowski spacetime, the subtractions performed using $H(x,x')$ can be
given a relatively simple 
mode-by-mode interpretation as ``vacuum subtractions''.
However, no such accident occurs in a general curved
spacetime. Indeed, in a general curved spacetime, $H(x,x')$ can be
defined only locally (i.e., for $x'$ close to $x$) and cannot be equal
to the expection value of $\varphi(x) \varphi(x')$ in {\em any} state.
Normal ordering with respect to some ``vacuum
state'' cannot yield a correct renormalization prescription in a
general curved spacetime---see remark (3) on P. 303 of \cite{hw} for a
formal proof of this statement---and 
there is no reasonable mode-by-mode interpretation of the
quantum field theoretic renormalization prescription.

Three concrete examples will serve to illustrate the above points. 
The first concerns the energy of a Klein-Gordon field,
eq.(\ref{kg}), 
in a $1+1$ dimensional static spacetime $(M, g_{ab})$
of spatial topology
$S^1$, with metric of the form
\begin{equation}
ds^{2} = - dt^2 + L^2 d\theta^2
\label{metric}
\end{equation}
where the range of the $\theta$-coordinate is $[0, 2 \pi)$. This
metric describes a flat universe with closed spatial sections of size
$2 \pi L$. As already mentioned above, the quantum field theoretic
renormalization prescription for defining the stress-energy tensor,
$T_{ab}$, of $\varphi$ involves a subtraction performed by using a
locally and covariantly constructed Hadamard distribution $H(x,x')$;
one subtracts from the ``point-split'' expression for $T_{ab}$ in terms 
of $\varphi(x) \varphi(x')$ a similar expression constructed using
$H(x,x')$ and then takes the coincidence limit.
However, since the spacetime metric (\ref{metric}) is locally flat,
for $x'$ near $x$ the locally constructed Hadamard distribution
$H(x,x')$ for the spacetime $(M, g_{ab})$ must be identical to
the corresponding locally constructed Hadamard distribution
for two-dimensional Minkowski spacetime. This means that
the renormalization prescription for $T_{ab}$ can be given the
interpretation of a ``vacuum subtraction'', but what is being
subtracted is {\em not} the vacuum energy of the modes of $\varphi$
that are actually present in $(M, g_{ab})$ but rather the vacuum
energy of the modes that hypothetically would have been present if the
spacetime were globally Minkowskian \cite{k}!
From the quantum field theoretic point of view, this prescription is
entirely natural, since the construction of the 
stress-energy tensor should be local
in the spacetime metric $g_{ab}$ and the quantum field $\varphi$,
and it therefore should not care about the global topology of $M$. But,
in terms of the individual globally defined modes, this prescription makes
no sense: Why should the appropriate subtractions be based upon the energies of
modes in some fictitious Minkowski spacetime rather than the energies
of the modes that are present in the actual universe $(M, g_{ab})$?

Using the quantum field theoretic renormalization methods, one finds
that the total energy, $E_C$, of the ground state\footnote{The
massless Klein-Gordon field does not actually have a ground state on
account of the presence of a spatally homogeneous mode which grows
linearly with time. This ``infrared divergence'' is not relevant to
any of our considerations.  All of our results rigorously apply to the
Klein-Gordon field of mass $m$ (which does have a ground state) in the
limit as $m \rightarrow 0$.} is given by \cite{bd}
\begin{equation}
E_C = - \frac{1}{12L} \, .
\label{Casimir}
\end{equation}
This result is in close analogy with the Casimir effect for a field
confined by conducting plates, which has been
verified experimentally \cite{bmm}. 
The negative value of $E_C$ can be understood
to arise from the fact that
there are, in a sense, fewer low energy modes of $\varphi$ in the
universe $(M, g_{ab})$ than in Minkowski spacetime, so the Minkowski
subtraction overcompensates for the vacuum energy of the modes in
$(M, g_{ab})$.

Let us now attempt to reproduce eq.(\ref{Casimir})
energy by applying quantum
mechanics to the low energy degrees of freedom of the quantum field
$\varphi$. If one decomposes $\varphi$ into its spatial
Fourier modes, one finds that the $n$th Fourier mode is precisely a
harmonic oscillator with frequency
\begin{equation}
\omega_n = \frac{n}{L} \, .
\end{equation}
Therefore, if each mode is put in its ground state but we only count
modes with frequency $\omega \lesssim 1/l_0$, we obtain the total energy
\begin{equation}
E_0 = \sum_{n=0}^{L/l_0} \frac{1}{2} \omega_n = \sum_{n=0}^{L/l_0} \frac{n}{2L}
\approx \frac{1}{4L} (L/l_0)^2 \, .
\label{E0}
\end{equation}
This disagrees with eq.~(\ref{Casimir}) not only in sign, but also by a 
factor of $(L/l_0)^2 \gg 1$. 

The above colossal value obtained for $E_0$ by this calculation is, of
course, very well known and comprises what is usually referred to as
the ``cosmological constant\footnote{The reason for this terminology
is that, in Minkowski spacetime, by Lorentz invariance, the expected
stress-energy in the vacuum state must be proportional to the
metric. Therefore, an absurdly large value of $<T_{ab}>$ would
correspond to the presence of an absurdly large value of the
cosmological constant in Einstein's equation. In fact, the insertion
of a high energy cutoff as we have done here breaks Lorentz invariance
and---since each individual mode contributes a traceless stress-energy
tensor---it is easy to see that the unrenormalized mode sum for
$T_{ab}$ corresponding to eq.(\ref{E0}) would not be proportional to
$g_{ab}$.  Nevertheless, we will use the conventional terminology in
referring to the enormous value of $E_0$ as the ``cosmological
constant problem''.}  problem''. However, we can attempt to ``fix''
the discrepancy between eqs.~(\ref{Casimir}) and (\ref{E0}) by
adjusting the zero of energy of the $n$th mode of the quantum field by
an amount $\epsilon_0 (n,L)$. 
As previously mentioned, the fact that one has to do a subtraction here
is not, by itself, necesarily worse than what was done 
in the quantum field theoretic
calculation, where a subtraction also was necessary. 
The revised formula for the total energy
of the ground state would then be
\begin{equation}
E'_0 = \sum_{n=0}^{L/l_0} \left[\frac{n}{2L} - \epsilon_0 (n,L)\right] \, .
\label{vacsub}
\end{equation}
Clearly, with unconstrained freedom on the choice of $\epsilon_0
(n,L)$, there is no difficulty in arranging for any answer that one
wishes to get for $E'_0$. Thus, one can, of course, make choices of
$\epsilon_0 (n,L)$ that yield agreement between the right sides of
eqs.~(\ref{Casimir}) and (\ref{vacsub}). The difficulty is that there
are infinitely many ways of doing this, and none of them appear to be
in any way natural. Indeed, the following would appear to be two very natural
conditions to impose on $\epsilon_0 (n,L)$. First, on a account of the
invariance of $\varphi$ under a scaling of the spacetime metric
(\ref{metric}), it would be natural to require $\epsilon_0 (n,L)$ to 
respect this scaling and therefore be of the form $\epsilon_0 (n,L) = f(n)/L$.
Second, since the mode labeled by the integer $Nn$ in a universe of
size $2 \pi N L$ is locally identical (up to normalization) to the mode
labeled by integer $n$ in a universe of size $2 \pi L$, it would be natural
to require $\epsilon_0 (n,L)$ to respect this fact by depending on $n$
and $L$ only in the form $n/L$. However, these two requirements would 
constrain $\epsilon_0 (n,L)$ to be of the form $cn/L$ for some constant $c$.
But this choice would then yield $E'_0 \approx (1-2c) L/4l_0^2$, which does
not agree with eq.~(\ref{Casimir}) for any choice of $c$. Thus, these
``natural'' requirements on $\epsilon_0 (n,L)$ are incompatible with
the Casimir effect. We do not believe that the Casimir effect can be
understood without invoking the holistic aspects of quantum field theory.

Our second example concerns the general problem of defining the
renormalized stress-energy tensor, $T_{ab}$, in a general,
time-dependent, globally hyperbolic, curved spacetime. We have just
argued that even in a static flat spacetime, the construction of
$T_{ab}$ cannot be understood without invoking the holistic aspects of
quantum field theory, so the situation in a general curved spacetime
cannot be better. However, we wish to point out that the general
situation is actually far worse, i.e., one must go to much greater lengths to
attempt to account for the stress-energy of quantum field by subtractions
performed on its individual degrees of freedom.

From the quantum field theoretic point of view, the calculation of
$T_{ab}$ in a general curved spacetime proceeds in much the same way
as indicated above for the Casimir effect. One again performs a
suitable subtraction using a locally and covariantly constructed
Hadamard distribution $H(x,x')$. The only additional complication is
that when spacetime curvature is present,
there is now a small amount of additional freedom in the
renormalization prescription, which allows one to modify the final
result for $T_{ab}$ by the addition of conserved local curvature
tensors of the correct scaling dimension \cite{w}. In 4-dimensions, there are
two such curvature tensors, so there is a two-parameter
freedom\footnote{If we were to consider a massive Klein-Gordon field, then we
also would have the additional freedom to modify the definition of
$T_{ab}$ by terms of the form $m^4 g_{ab}$ and $m^2 G_{ab}$.} in the
definition of $T_{ab}$.

Let us now try to construct $T_{ab}$ by applying quantum mechanics to
the low energy modes of $\varphi$, without invoking any holistic
aspects of quantum field theory. We immediately face a serious problem
in that it is far from clear how to even define the ``modes'' of
$\varphi$: The decomposition of a quantum field into modes requires a
definition of ``positive frequency'', but there is no natural
positive/negative frequency decomposition of $\varphi$ in a
non-stationary spacetime. Nevertheless, we can proceed by making some
arbitrary choice of ``positive frequency'', corresponding to some
arbitrary choice of ``vacuum state'' $|0 \rangle$. We can then write
down a mode sum formula for $\langle 0|T_{ab}|0 \rangle$ analogous to
eq.~(\ref{E0}). As in eq.~(\ref{E0}), we will obtain an enormous value
for $\langle 0|T_{ab}|0 \rangle$, so some subtractions are needed. But
it is hard to imagine that there could be any natural rule on what to
subtract from each mode that would yield agreement with the quantum
field theoretic expression. In particular, as already noted above,
subtraction of the entire vacuum stress-energy---i.e., normal
ordering---is incompatible with the quantum field theoretic
prescription. Furthermore, taking into account the known dependence of
$H(x,x')$ on the spacetime metric, it is not difficult to see that in
$D$ dimensions, the quantum field theoretic prescription for defining
$T_{ab}$ involves the local subtraction of terms that depend upon the
derivatives of the metric up to $D$th order. This means that if we
consider a two-dimensional spacetime with metric of the form
(\ref{metric}) but with $L$ now allowed to depend upon $t$, then in
order to reproduce the quantum field theoretic prediction for energy,
it would be necessary for the ``vacuum energy subtraction''
$\epsilon_0$ to depend not only on $n$ and $L$ but also on $dL/dt$ and
$d^2L/dt^2$---even though the Hamiltonian for the individual modes depends 
only on $n$ and $L$. Again, we do not believe that it is possible to sensibly
derive the quantum field theoretic prediction for $T_{ab}$ without
invoking the holistic nature of quantum field theory.

Our final example concerns anomalies. One of the most surprising
aspects of quantum field theory is that certain relations involving
the field equations that are manifestly true in classical field theory
cannot be satisfied in quantum field theory. For example, in the case
of a Klein-Gordon field $\varphi$ in 4-dimensional curved spacetimes, 
although
eq.~(\ref{kg}) of course holds, it is impossible to define the renormalizations
so as to satisfy both of the following relations~\cite{hw2}
\begin{equation}
\varphi \nabla^a \nabla_a \varphi = 0
\label{a1}
\end{equation}
\begin{equation}
\nabla_b \varphi \nabla^a \nabla_a \varphi = 0 \, .
\label{a2}
\end{equation}
(In the case of a conformally invariant field, the similar inability
to simultaneously impose analogs of eqs.~(\ref{a1}) and (\ref{a2}) is
responsible for the existence of a trace anomaly in the stress-energy
tensor of that field.) From the quantum field theoretic point of view,
the above anomaly arises because in a general curved spacetime it is
impossible to locally construct a Hadamard distribution $H(x,x')$ that
satisfies the Klein-Gordon equation in both $x$ and $x'$
\cite{w2}. Consequently, the quantum field theoretic subtraction
procedure fails to fully respect the Klein-Gordon equation.

Any attempt to reproduce the above anomaly by applying quantum
mechanics to the low energy modes of $\varphi$ would have to be truly
bizarre. Each term in the mode sum formulas for the left sides of
eqs.~(\ref{a1}) and (\ref{a2}) would vanish, since the individual
modes themselves do not suffer any anomalies. Yet, one would
nevertheless have to do some ``subtraction'' to obtain agreement with
the quantum field theoretic prediction. Again, we do not believe that
the existence of anomalies can be understood without invoking the
holistic nature of quantum field theory.

As already mentioned above, the absurdly large value obtained for the
stress-energy of a quantum field when computed by applying quantum
mechanics without subtractions to the low energy modes of the field is
usually referred to as the ``cosmological constant problem''.  In
4~dimensions, a calculation similar to that leading to eq.~(\ref{E0})
above would yield an expected energy density of order $1/l_0^4$, where
$1/l_0 \sim 10^{19}$ GeV if $l_0$ is assumed to be of order the Planck
length. By contrast, the actual energy density of our universe is only of
order $[\sim 10^{-12} \rm{GeV}]^4$. If one were to view a quantum
field as a collection of independent degrees of freedom that know
nothing about each other, then it is hard to imagine how---even
allowing for reasonable renormalization ``subtractions'' of the
individual modes---cancellations of this magnitude could occur.
Furthermore, in the absence of some exact, unbroken symmetry such as
supersymmetry, it is equally hard to imagine how cancellations of this
magnitude could occur between different fields. Thus, the enormous discrepancy
between the naive mode-sum calculation and the observed energy density
is therefore generally viewed as a very serious ``problem''. We do not
share this view. As we have argued above, there are many aspects of
the theory of a quantum field that simply cannot be understood by
viewing its low energy degrees of freedom as being independent. The
mode sum calculations like the one leading to eq.~(\ref{E0}) do not
properly take into account the holistic aspects of quantum field
theory.  In our
view, it would be more fruitful to simply accept the holistic aspects
of quantum field theory rather than search for models where---by virtue
of miraculous cancellations---the holistic aspects do not need to be
invoked\footnote{Of course, it remains a very significant puzzle as to why
quantum field theory possesses holistic aspects, i.e., how they arise
from the more fundamental, underlying theory. However, it is
likely that we will need a much deeper understanding of the
underlying theory in order to account for this.}.

If one accepts the holistic aspects of quantum field theory, there is
still a ``cosmological constant problem'', but it is rather different
than the usual formulation of it.  The puzzle is not, ``Why is the
observed energy density of the universe so small?''  This is only a
puzzle if one believes that it should be correct to calculate the
stress-energy of a quantum field by treating its low energy modes as completely
independent degrees of freedom that know nothing about each other, in
which case implausible cancellations would be required. Rather, the puzzle is,
``Why is the cosmological constant so {\em large}?''  Quantum field
theory predicts that the stress-energy tensor of a free quantum field in
an adiabatic vacuum state in a slowly expanding 4-dimensional
universe should be of
order of $L^{-4}$, where $L$ denotes the size and/or radius of
curvature of the universe. For our universe, $1/L$ would be of order
$\sim 10^{-42}$ GeV. But observations of type Ia supernovae and the
cosmic microwave background strongly suggest that, at the present
time, the dominant component of stress-energy in the universe is
smoothly distributed (i.e., not clustered with galaxies) and has
negative pressure. The energy density of this so-called ``dark
energy'' is thus $[\sim 10^{-12} \rm{GeV}]^4$, i.e. roughly the geometric
mean of the unsubtracted mode sum and quantum field theoretic predictions
for vacuum energy density. It is, of course, not presently known
whether this dark energy corresponds to the vacuum energy of some field, 
the potential
energy of some field, some other form of matter, 
or simply corresponds to the presence of a cosmological constant term in
Einstein's equation. 
In any case, however,
it is seems very difficult to account for its energy scale. 
This is a true puzzle. We do not
have any new proposals to make here concerning the nature of dark
energy. However, if dark energy does
correspond to vacuum energy of an interacting quantum field, it is
our view that its properties will be understood only by fully taking
into account the holistic nature of quantum field theory.

\medskip

This research was supported in part by NSF grant
PHY00-90138 to the University of Chicago.

\end{document}